\documentclass[%
preprint,
superscriptaddress,
amsmath,amssymb,
 aps,
prf,
]{revtex4-2}
\usepackage{graphicx}
\usepackage{bm}
\usepackage{amsmath}
\usepackage{bm}
\begin{document}

\title{Capillary nanowaves on surfactant-laden liquid films with surface viscosity and elasticity}
\author{Yixin Zhang}
\email{Y.Zhang-11@utwente.nl}
  \affiliation{Physics of Fluids Group, Max Planck Center Twente for Complex Fluid Dynamics and J. M. Burgers Centre for Fluid Dynamics, University of Twente, P.O. Box 217, 7500 AE Enschede, The Netherlands}
  \author{Zijing Ding}
  \affiliation{School of Energy Science and Engineering, Harbin Institute of Technology, Harbin, China}
\begin{abstract}
Thermal motions of molecules can generate nanowaves on the free surface of a liquid film. As nanofilms are susceptible to the contamination of surfactants, this work investigates the effects of surfactants on dynamics of nanowaves on a bounded film with a finite depth, using both molecular dynamics simulations and analytical theories. In molecular simulations, a bead-spring model is adopted to simulate surfactants, where beads are connected by the finite extensive nonlinear elastic potentials. Fourier transforms of the film surface profiles $h(x,t)$ extracted from molecular simulations are performed to obtain the static spectrum $|h_q|_{\mathrm{rms}}$ and temporal correlations of surface modes $<h_q(0)h_q^*(t)>$. It is shown that the spectral amplitude is increased for the contaminated liquid surface compared to the clean surface because surfactants can decrease surface tension. A higher concentration of surfactants on the surface not only decreases the surface tension but also causes elastic energy to the free surface, as the scaling of spectral amplitude with wavenumbers changes from $|h_q|_{\mathrm{rms}}\sim q^{-1}$ to $|h_q|_{\mathrm{rms}}\sim q^{-2}$ for modes with large wavenumbers. Regarding the temporal correlations of surface modes, it is observed that the presence of surfactants leads to a slower decay, which, however, cannot be predicted by only considering the decreased surface tension. Based on the Boussinesq–Scriven model for surface viscosity, a linear stability analysis of Stokes flow for films with arbitrary depth is conducted and the obtained dispersion relation considering surface viscosity can justify the simulation results.
\end{abstract}
\maketitle 
\newpage 
\section{introduction}
Surfactants are widely used in a number of industrial applications such as foams and emulsions\,\cite{sadoc2013foams}, detergents, inks\,\cite{deng2018surfactant}, and oil recovery\,\cite{negin2017most}. The presence of surfactants can significantly alter the behaviors of liquid surfaces. For example, surfactants in water may lead to a Marangoni stress on the bubble surface that can slow the rising of bubbles\,\cite{takagi2011surfactant}. Thus, it is of great interest and importance to study the effects of surfactants on various kinds of interfacial fluid dynamics elaborately.

Thermal capillary waves (TCW) on a free liquid surface are waves spontaneously excited by thermal motions of molecules. These waves have been used in experiments to measure fluid properties such as surface tension\,\cite{ki2003} and viscoelasticity\,\cite{ji2007}. These waves are also critical to the instability of nanofilms\,\cite{gr2006,zhang2019,vrij1966possible,vrij1968rupture}, coalescence of nanodroplets\,\cite{perumanath2019droplet} and bubbles, and rupture of nanojets\,\cite{zhao2019revisiting}. The roughness on a liquid surface created by TCW is usually on the scale of nanometres, but micrometer roughness can also be obtained and thus observed optically using ultra-law surface tension mixtures ($\gamma \sim 10^{-6}$ N/m)\,\cite{aa2004}, since thermal roughness is usually proportional to thermal length $\sqrt{k_BT/\gamma}$ ($k_B$, $T$, and $\gamma$ are the Boltzmann constant, temperature, and surface tension respectively). It is worth mentioning that the amplitude of TCW will diverge with the increasing system size if only considering surface tension on the free surface. The introduction of gravity or a certain binding potential, however, can remedy this problem\,\cite{rowlinson2013molecular}.

At thermal equilibrium, the amplitude of TCW, namely the static spectrum, can be described by the famous capillary wave theory\,\cite{buff1965interfacial,ma2017}. Recently, the capillary wave theory has been extended using a Langevin equation to describe the transient dynamics of non-equilibrium TCW and their approach to thermal equilibrium\,\cite{zhang2020thermal}. When in thermal equilibrium, it is well known that the temporal correlations of overdamped surface modes of TCW show an exponential decay\,\cite{he2007,zhang2021relaxation}, with a decay rate given by the dispersion relation of the system, which provides the basis for measuring the system's property using TCW\,\cite{ki2003,ji2007}. This technique has the advantage of being noninvasive since no external forces are applied. As the surface of a nanometric liquid film can often be contaminated by surface-active agents such as surfactants\,\cite{lohse2015surface,maali2017viscoelastic}, it is crucial to understand how surfactants modify the properties of TCW. 

Perhaps the most known effect of surfactants is their ability to decrease surface tension. Thus, adding surfactants to a liquid surface may enhance the amplitude of thermal waves and roughness. Apart from decreasing surface tension, surfactants may form a monolayer on the liquid surface and lead to the elastic properties of the liquid surface\,\cite{langevin2014rheology,jiang2022thin}. For example, the stability of emulsions, found in mixtures of water, oil, and amphiphilic surfactants, may be influenced by the bending rigidity ($\kappa$) of the surfactant monolayers\,\cite{helfrich1978steric,laradji2000elastic}.  The bending rigidity of emulsions is on the order of $\kappa \sim k_B T$ and may overwhelm the effects of surface tension\,\cite{laradji2000elastic}. It is thus interesting to see how the bending rigidity affects thermal capillary waves. In fact, a recent experiment used the static spectra of TCW in an oil-water-surfactant system to measure the bending rigidity of the water-oil interface of droplets\,\cite{bolognesi2016mechanical}.

The adsorption of surfactants on a fluid surface may also increase that surface's resistance to deformations and lead to a new surface property called surface viscosity\,\cite{scriven1960marangoni}. The importance of surface viscosity on free surface flows has been generally acknowledged now. For example, Joye \textit{et al.}\,\cite{joye1994asymmetric} experimentally showed that high surface viscosity can prevent asymmetric drainage of thin films in foams. Ponce-Torres \textit{et al.}\,\cite{ponce2017influence} found surface viscosity is responsible for the accumulation of surfactants in the satellite droplets formed by the breakup of a large surfactant-laden droplet. Surface viscosity can also change the instability and pinch-off profile of viscous threads\,\cite{martinez2018temporal, wee2020effects}. For thin liquid films destabilized by the disjoining pressure, surface viscosity is able to decrease the growth of perturbations\,\cite{edwards1995instability,choudhury2020enhanced}. Shen \textit{et al.}\,\cite{shen2018capillary} theoretically showed surface viscosity contributes an overall damping effect to the amplitude of the capillary waves on films with infinite depth. However, there are very few studies about the effects of surface viscosity on TCW. Given the importance of surface viscosity on free surface flows, seeking accurate methods to measure surface viscosity is necessary. Though much progress has been made, it is still debatable how to obtain reliable values of surface viscosity, since Marangoni effects and surface viscosity can often coexist during measurements\,\cite{elfring2016surface}. Experimental measurements of surface viscosity have reported values that are orders of magnitude apart and are controversial\,\cite{stevenson2005remarks}.

As experiments at the nanoscale are difficult to perform due to the small temporal and spatial scales, molecular dynamics simulations (MD) become a valuable tool to investigate nanoflows and can act as virtual experiments to validate newly developed theories. In terms of nanoflows involving free surfaces, there are increasing molecular studies for nanodroplets, nanobubbles\,\cite{dockar2018mechanical}, nanofilms\,\cite{zhang2019} and nanojets\,\cite{zhao2019revisiting}. MD studies of TCW have been carried out for a variety of problems including TCW for free liquid films\,\cite{delgado2008hydrodynamics},  TCW for films on no-slip\,\cite{wi2010} and anisotropic-slip\,\cite{zhang2021relaxation} substrates, to name a few. The introduction of surfactants to TCW in MD has been implemented in several studies\,\cite{laradji2000elastic,rekvig2004chain} but previous works only considered the effects of surfactants on surface tension and bending rigidity of liquid surface by examining the static spectrum. The dynamics of TCW such as the relaxation of TCW correlations associated with surface viscosity and elasticity have not been studied.

In this work, MD simulations are used to study the effects of surfactants on the overdamped thermal capillary waves for liquid films bounded on substrates. A bead-spring model is adopted to simulate surfactants, where beads are connected by the finite extensive nonlinear elastic potentials. The surface concentrations of surfactants are varied to see how they change the behaviors of the liquid interface. We obtain surface modes of surface waves from MD simulations and calculate their static spectra $|h_q|_{\mathrm{rms}}$ and temporal correlations $<h_q(0)h_q^*(t)>$. The static spectrum is used to infer bending rigidity and surface tension. The Boussinesq–Scriven model for the surface viscosity is adopted and we perform a linear stability analysis of a liquid film with arbitrary depth in Stokes limits and obtain the corresponding dispersion relation for the first time, which is validated against the MD temporal correlations. 

This paper is organized as follows. In Sec.\,\ref{sec2}, we formulate the problem that we are going to solve and derive the dispersion relation for films with any depth and surface viscosity. In Sec.\,\ref{sec3}, the MD model of nanoscale liquid films with surfactants is introduced. Sec.\,\ref{sec4}, shows the comparison between MD and analytical theories. We conclude our findings and outline future directions of research in Sec.\,\ref{sec5}.
\section{Mathematical modeling}\label{sec2}
\begin{figure}[h] 
\includegraphics[width=\linewidth]{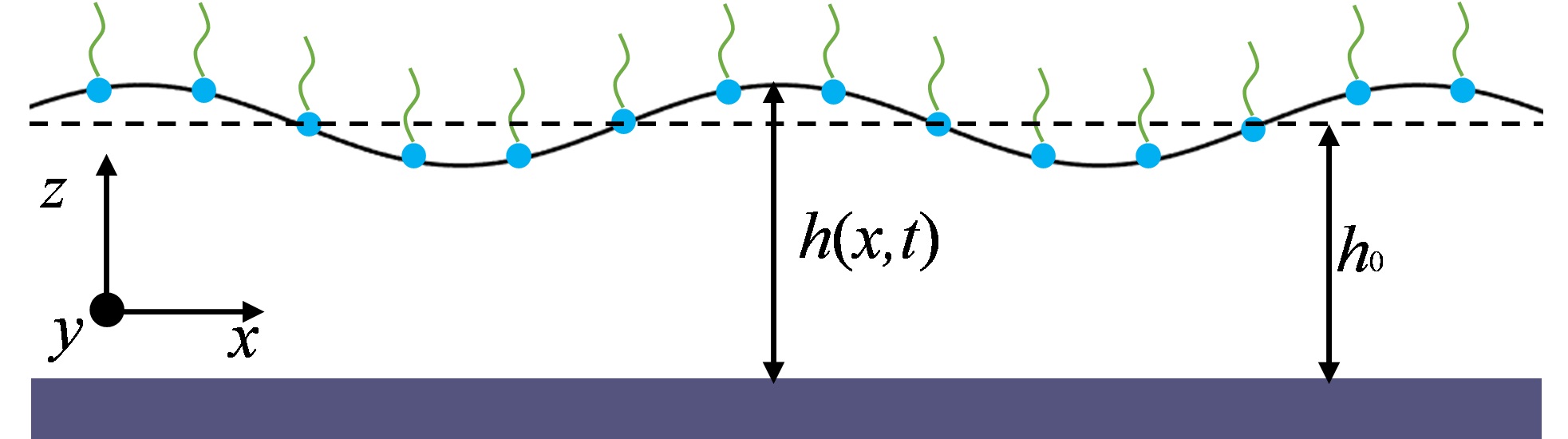}
\caption{\label{fig1} Sketch of a (quasi-two dimensional) surfactant-laden liquid film on a plate. The $h_0$ is the initial film thickness and $h=h(x,t)$ is the film thickness under spontaneous perturbations due to thermal fluctuations. The film has a small depth $L_y$ in the $y$ direction (into the page).}
\end{figure}

As shown in Fig.~\ref{fig1}, we consider a Newtonian liquid film on a solid surface. The free surface of the liquid film is covered with insoluble surfactants. Initially, the film has a size $\left(L_x,L_y,h_0\right)$ and is quasi-two dimensional (2D) by letting $L_x\gg L_y$. Due to thermal motions of molecules, the film surface is fluctuating weakly around $h_0$ and the instantaneous surface profile is given by $h(x,t)$.
\subsection{Static spectrum and temporal correlations for surfactant-laden liquid films}
The static spectrum of capillary waves on a clean film can be determined by the equipartition theorem; this forms the basis of classical capillary wave theory\,\cite{ma2017}. For a film contaminated with surfactants, the static spectrum has to be modified to consider the contributions of the elastic energy arising from the formed monolayer of surfactants. In terms of a quasi-2D film under small perturbations in Fig.~\ref{fig1}, the extra free energy $f$ due to the change of surface area is\,\cite{laradji2000elastic,helfrich1978steric}
\begin{equation}
f\approx \frac{L_y}{2}\gamma \int{{{\left( \frac{\partial h}{\partial x} \right)}^{2}}}dx+\frac{L_y}{2}\kappa \int{{{\left( \frac{\partial^2 h}{\partial x^2}\right)}^{2}}}dx.
\end{equation}
Note that the effect of disjoining pressure is ignored as we are going to study a stable film where the effect of disjoining pressure is weak. As you will see, in our MD simulations, the thickness of the film is much larger than the cut-off distance so that the effect of disjoining pressure is safely neglected. Let us define the Fourier transform of $\delta h=h(x)-h_0$ as $h_q=\int{\delta h{{e}^{-iqx}}dx}$. From Parseval's theorem, we can express Eq.~(1) in terms of Fourier modes:
\begin{equation}
f=\frac{1}{2}\frac{{{L}_{y}}}{{{L}_{x}}}\gamma \sum{{{q}^{2}}{{\left| h_q \right|}^{2}}}+\frac{1}{2}\frac{{{L}_{y}}}{{{L}_{x}}}\kappa \sum{{{q}^{4}}{{\left| h_q \right|}^{2}}}.
\end{equation}
As each summand appears quadratically, it has the same energy $\frac{1}{2}k_BT$, from the equipartition theorem, so that
\begin{equation}
\frac{1}{2}{{k}_{B}}T=\frac{{{L}_{y}}}{{{L}_{x}}}\left( \frac{1}{2}\gamma {{q}^{2}}+\frac{1}{2}\kappa {{q}^{4}}\right){{\left| h_q \right|}^{2}}.
\end{equation}
Thus, the static spectrum for a surfactant-laden film is derived:
\begin{equation}
 S_s=\sqrt{\left\langle\left| h_q\right|^2 \right\rangle}=\sqrt{\frac{{{L}_{x}}}{{{L}_{y}}}\frac{{{k}_{B}}T}{\gamma {{q}^{2}}+\kappa q^4}}.
\end{equation}
Here $\left\langle ...\right\rangle$ represents an ensemble average.

In thermal equilibrium, the temporal correlations of overdamped capillary waves usually decay exponentially to zero \cite{ki2003,wi2010}
\begin{equation}\label{eqe}
C_{h_qh_q^*}=\frac{\left\langle  h_q\left(q,t\right){{h_q}^{*}}\left(q,t'\right) \right\rangle}{S_s^2}={{e}^{\Omega \left(q\right)\left| t-t' \right|}},
\end{equation}
where the asterisk denotes conjugate values. The decay rate $\Omega$ ($\Omega<0$) is given by the dispersion relation of the system (temporal growth rate of a surface mode), which is derived by a linear stability analysis (LSA) in the next subsection. 

\subsection{LSA of Stokes flow for films with arbitrary depth and surface viscosity }    
The liquid is assumed to be incompressible
\begin{equation}
\nabla \cdot \mathbf{u}=0,
\end{equation}
where $\mathbf{u}=(u_x,u_z)$, and $u_x,u_z$ are the velocities in $x$ and $z$ directions, respectively. The momentum equation with the assumption of Stokes flow (the Reynolds number is small), is as follows
\begin{equation}
\mu\nabla^2\mathbf{u}=\nabla p,
\end{equation} 
where $\mu$ and $p$ are the viscosity and pressure of the liquid. Note that transient inertia may come into play depending on the frequency of the perturbations, which leads to a critical wavenumber below which the correlations are underdamped instead of being overdamped\,\citep{delgado2008hydrodynamics}. However, as one can see from our following molecular simulations, all measured correlations are overdamped.

For the dynamic boundary condition at the liquid-vapour interface $z=h(x,t)$, we have the leading-order Boussinesq–Scriven model for one-dimensional surface~\cite{shen2018capillary}
\begin{equation}
\mathbf{n}\cdot\bm{\tau}=\nabla_s \tilde{\gamma}-\tilde{\gamma}\left( \nabla_s\cdot\mathbf{n}\right)\mathbf{n}+\mathbf{f},
\end{equation}
where $\bm{\tau}$ is the hydrodynamic stress tensor, $
\tau_{ij}=-p\delta_{ij}+\mu\left(\partial u_i/\partial x_j+\partial u_j/\partial x_i\right)$. The $\tilde{\gamma}$ is the surface tension modified by the surface viscosity, $\tilde{\gamma}=\gamma+\left(\mu_d+\mu_s\right)\nabla_s\cdot\mathbf{u}$. $\nabla_s$ is the surface gradient. The dilatational ($\mu_d$) and shear ($\mu_s$) components of surface viscosity
occur in additive pairs in a 2D system and will hereafter be written in terms of a single parameter, $\eta=\mu_d+\mu_s$ in the rest of this paper\,\cite{edwards1995instability,choudhury2020enhanced}. The $\mathbf{f}=\kappa \frac{\partial^4 h}{\partial x^4}\mathbf{n}$ is the elastic pressure\,\cite{young2017long}. Finally, $\mathbf{n}$ is the outward normal to the free surface:
\begin{equation}
\mathbf{n}=\frac{\left( -\partial h/\partial x,1 \right)}{\sqrt{1+{{\left( \partial h/\partial x \right)}^{2}}}}.
\end{equation}
Under the assumption of small perturbations (${\partial h}/{\partial x}\ll 1$) the dynamic boundary condition is reduced to (in the normal direction):
\begin{equation}
-p+2\mu\frac{\partial u_z}{\partial z}=\gamma \frac{\partial^2 h}{\partial x^2}+\kappa \frac{\partial^4 h}{\partial x^4},
\end{equation}
and in the tangential directions to the surface
\begin{equation}
\mu\left(\frac{\partial u_x}{\partial z}+\frac{\partial u_z}{\partial x}\right)=\eta\frac{\partial^2 u_x}{\partial x^2},
\end{equation}
where we have assumed surface tension (concentration of surfactants) is uniform along the surface so that the ${\partial \gamma}/{\partial x}$ term is negligible. This is because, at the nanoscale, surface diffusion can counter advective effects and result in a uniform surfactant distribution\,\citep{timmermans2002effect}. This is also directly found in our MD simulations shown in the appendix.

The kinematic condition at the free surface is given by
\begin{equation}
u_z=\frac{\partial h}{\partial t}+u_x\frac{\partial h}{\partial x}.
\end{equation}
At $z=0$, the no penetration condition and Navier's slip condition are,
\begin{align}
u_z&=0,\\
u_x&=\ell\frac{\partial u_x}{\partial z}.
\end{align} 

Equations (6-14) are linearised using normal modes
 \begin{equation}
 \left(u_x,u_z,p,h-h_0\right)=\left(\tilde{u}_x,\tilde{u}_z,\tilde{p},\tilde{h}\right){{e}^{\Omega t+i{{q}}x}}.
 \end{equation} 
With those, the expression for the single variable $\tilde{u}_z$ can be obtained from Eq.~(6) and Eq.~(7) as:
\begin{equation}
\frac{{{d}^{4}}\tilde{u}_z}{d{{z}^{4}}}-2{{q}^{2}}\frac{{{d}^{2}}\tilde{u}_z}{d{{z}^{2}}}+{{q}^{4}}\tilde{u}_z=0.
\end{equation}
The general solution for $\tilde{u}_z$ is thus\,\cite{he2007}
\begin{equation}
\tilde{u}_z={{C}_{1}}\cosh \left( qz \right)+{{C}_{2}}\sinh \left( qz \right)+{{C}_{3}}qz\cosh \left( qz \right)+{{C}_{4}}qz\sinh \left( qz \right),
\end{equation}
and one can also obtain the solution for $\tilde{u}_x$, and $\tilde{p}$: 
\begin{align}
 & \tilde{u}_x=\frac{i}{q}\frac{d\tilde{u_z}}{dz}, \\ 
 & \tilde{p}=2\mu q\left[ {{C}_{3}}\cosh \left( qz \right)+{{C}_{4}}\sinh \left( qz \right) \right].
\end{align}
Here, $C_1-C_4$ are four coefficients to be determined by the boundary conditions Eqs.~(10-14), whose linearised forms are
\begin{subequations}
\begin{equation}
-\tilde{p}+2\mu\frac{d\tilde{u}_z}{d z}=-\gamma q^2\tilde{h}-\kappa q^4\tilde{h},\quad \mbox{at\ }\quad z=h
\end{equation}
\begin{equation}
\mu\left(\frac{d\tilde{u}_x}{dz}+iq_x\tilde{u}_z\right)=-\eta q^2 \tilde{u}_x,\quad \mbox{at\ }\quad z=h
\end{equation}
\begin{equation}
\Omega=\frac{\tilde{u}_z}{\tilde{h}},\quad \mbox{at\ }\quad z=h
\end{equation}
\begin{equation}
\tilde{u}_z=0,\quad \mbox{at\ }\quad z=0
\end{equation}
\begin{equation}
\tilde{u}_x=\ell\frac{d\tilde{u}_x}{dz}.\quad \mbox{at\ }\quad z=0
\end{equation}
\end{subequations}

Substituting Eqs.~(17-19) into Eqs.~(20) leads to the required dispersion relation
\begin{gather} 
\Omega =-\frac{\gamma {{q}^{2}}+\kappa {{q}^{4}}}{\mu q}\frac{B_1}{B_2}, \nonumber\\ 
\begin{align}\label{eq:disp_rela}
  B_1= &\mu\left[\sinh \left( 2qh_0 \right)-2qh_0+4ql{{\sinh }^{2}}\left( qh_0 \right)\right]\nonumber\\
 &+\left\{ {{\sinh }^{2}}\left( qh_0 \right)-{{q}^{2}}{h_0^{2}}+ql\left[ \sinh \left( 2qh_0 \right)-2qh_0 \right] \right\}{{\eta }}q, \nonumber\\
 B_2=&\mu\left\{ 4{{q}^{2}}{h_0^{2}}+4{{\cosh }^{2}}\left( qh_0 \right)+4ql\left[ 2qh_0+\sinh \left( 2qh_0 \right) \right]\right\}\nonumber\\
 &+\left[ 2qh_0+\sinh \left( 2qh_0 \right)+4ql{{\cosh }^{2}}\left( qh_0 \right) \right]{{\eta }}q.  
\end{align}
\end{gather}
\section{Molecular dynamics simulations}\label{sec3}
\begin{figure}[h]
\includegraphics [width=\linewidth]{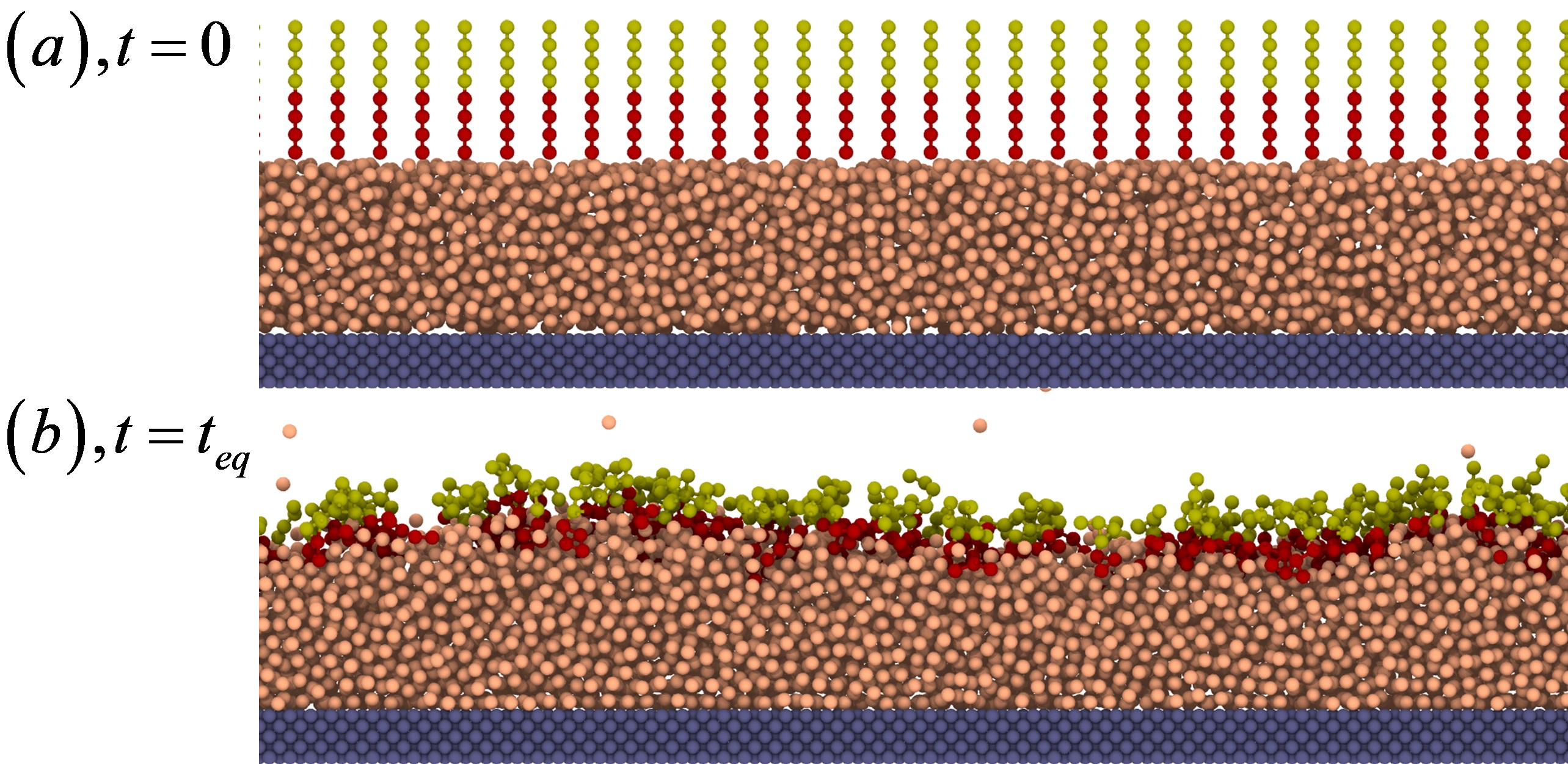}
\caption{\label{fig2} Snapshots of MD simulations of a surfactant-laden liquid film on a substrate. (a) Initial setting. The fluid atoms are colored in orange and the solid atoms are navy blue. The surfactant is made of chained beads with a hydrophobic tail (in bronze) and a hydrophilic head (in red). (b) Equilibrium configurations of a surfactant-laden interface.}
\end{figure}

Molecular dynamics simulations are used to simulate surfactant-laden liquid films on substrates. These simulations are performed using the open-source code LAMMPS\,\cite{pl1995}. The molecular system contains fluid atoms (liquid and its vapor), surfactant molecules, and solid atoms, as shown in Fig.~\ref{fig2}(a). A surfactant molecule $H_mT_n$ consists of $m+n$ atoms and is amphiphilic with $m$ hydrophilic atoms (the head group) and $n$ hydrophobic atoms (the tail group).

The non-bonded intermolecular potentials $U$ between $i$-type atoms and $j$-type atoms are simulated with the standard Lennard-Jones (LJ) 12-6 potential:
\begin{equation}
  U({{r}_{ij}}) =
    \begin{cases}
      4\varepsilon_{ij} \left[ {{\left( \frac{\sigma_{ij} }{{{r}_{ij}}} \right)}^{12}}-{{\left( \frac{\sigma_{ij} }{{{r}_{ij}}} \right)}^{6}} \right] & \text{if} \,\,\,{{r}_{ij}}\le {{r}_{c,ij}},\\
      0 & \text{if}\,\,\, {{r}_{ij}}>{{r}_{c,ij}},\\
    \end{cases}       
\end{equation}
where $r_{ij},\varepsilon_{ij},\sigma_{ij}$ and ${r}_{c,ij}$ are the pairwise distance, energy parameter, length parameter, and cut-off distance respectively. The complete lists of parameters among the liquid (l), solid (s), head group of surfactants (h), and tail group of surfactants (t) are given in Table \ref{tab1}.
\begin{table}[h!]
\centering
 \begin{tabular}{c c c c c } 
 \hline
 Atom type & Atom type  & $\varepsilon_{ij}/\varepsilon_{ll}$ & $\sigma_{ij}/\sigma_{ll}$  & $r_{c,ij}/\sigma_{ll}$ \\ 
 \hline
  L & L & 1 & 1 & 5.5    \\ 
 \hline
  L &  S & 0.65 & 0.8 & 5.5  \\
 \hline
  L & H & 0.80 & 1 & 5.5 \\
  \hline
  L & T & 1 & 1 &  $2^{1/6}$     \\ 
  \hline
  H & H & 0.5 & 1 & 5.5    \\ 
  \hline
  H & T & 1 & 1 & $2^{1/6}$    \\ 
  \hline
  H & S & 1 & 1 & $2^{1/6}$    \\ 
    \hline
  T & T & 0.35 & 1 & $2^{1/6}$    \\ 
    \hline
  T & S & 1 & 1 & $2^{1/6}$    \\ 
 \hline
\end{tabular}
\caption{{\label{tab1}}Interaction parameters among the liquid (L), solid (S), head group of surfactants (H), and tail group of surfactants (T).}
\end{table}

The liquid is simulated as argon whose $\varepsilon_{ll}$, $\sigma_{ll}$, and atomic mass are $1.67\times{10^{-21} }$ J, 0.34 nm, and $6.63\times{10^{-26}}$ kg, respectively.  The temperature of this system is kept at $T=85$ K or $T^*=0.7 \varepsilon_{ll} /k_B$ (* henceforth denotes LJ units and $k_B$ is the Boltzmann constant) using the Nos\'{e}-Hoover thermostat. At this temperature, the mass density of liquid argon is $1.40\times{10^{3}}$ kg/m\textsuperscript{3} and number density $n_{l}^*=0.83/\sigma_{ll}^3$. The cut-off distance for liquid-liquid interactions, beyond which the intermolecular interactions are omitted, is chosen as ${r_{c,ll}}^*=5.5 \sigma_{ll}$. Under above parameters and conditions, the surface tension of the clean liquid is $\gamma_0 = 1.52\times{10^{-2}} $ N/m obtained by the mechanical route\,\cite{kirkwood1949statistical}. The dynamic viscosity of liquid is $\mu=  2.87\times{10^{-4}}$ kg/(ms) calculated by the Green-Kubo relation. 

The substrate is platinum with a face-centred cubic (fcc) structure, and we use its isotropic $\left\langle 100\right\rangle$ surface to contact the fluid\,\cite{zhang2021relaxation}. The platinum mass density is $21.45\times{10^3}$ kg/m\textsuperscript{3} with an atomic mass of $3.24\times{10^{-25}}$ kg. The solid substrate is rigid in MD simulations. The liquid-solid interactions are modelled by the same 12-6 LJ potential with $\varepsilon_{ls} =C\varepsilon_{ll}$ and $\sigma_{ls}=0.8\sigma_{ll}$. One may vary $C$ to obtain different amounts of slip. Here we choose $C=0.65$ so that there is nearly no slip between liquid and solid, since the effects of slip on capillary wave dynamics have been examined elaborately in our previous works\,\cite{zhang2020thermal,zhang2021relaxation}. 

To simulate surfactants, a coarse-grained model called the `bead-spring' model is adopted\,\cite{kremer1990dynamics,zhang2022fate}. The surfactant molecule $H_mT_n$ consists of $m+n$ atoms connected by the finite extensible nonlinear elastic (FENE) potential\,\cite{kremer1990dynamics}
\begin{equation}
U=-0.5KR_0^2\mathrm{ln}\left[1-\left(\frac{r}{R_0}\right)^2\right].
\end{equation} 
Here $K^*=12 \varepsilon/\sigma^2$ and $R_0^*=1.4\sigma$ are used. The tail group of a surfactant molecule is hydrophobic and this is achieved by setting the $r^*_{c,lt}=2^{1/6}\sigma_{ll}$ so that interactions between liquid and the tail group are purely repulsive. In our simulations, the type of surfactant $H_4T_4$ is used. 

The initial dimensions of the liquid film $(L_x, L_y, h_0)$ in Fig.~\ref{fig1}(a) are chosen as $L_x=31.4$ nm, $ L_y = 3.14$ nm, and $h_0=3.14$ nm. The solid substrate has the same lateral size as that of the liquid film and has a thickness $h_s=0.78$ nm. As $L_y \ll L_x$, the system is quasi-2D.  Periodic boundary conditions are applied in both the $x$ and $y$ directions of the system whilst vapour particles are reflected specularly in the $z$ direction at the top boundary of the system.

In one single simulation of the surfactant-laden film, the simulation with the initial setting shown in Fig.~\ref{fig2}(a) is run for 20 ns with a timestep 8.57 fs to reach its equilibrium state, see Fig.~\ref{fig2}(b). After the surface has reached equilibrium, the simulation is run to output the positions of atoms every 2000 steps. The free surface position is defined as the usual equimolar surface and the way to extract the surface profile $h(x,t)$ from positions of atoms in MD simulations is detailed in our previous work\,\cite{zhang2019}. After obtaining $h(x,t)$, Fourier transforms are performed to obtain the amplitude of surface modes $h(q,t)$.
\section{Results and Discussions}\label{sec4}
In this section, we present and discuss the results of MD simulations and their comparison to analytical solutions.
\subsection{Enhanced static spectrum, surface roughness and elasticity}
\begin{figure}[h]
\includegraphics [width=0.5\linewidth]{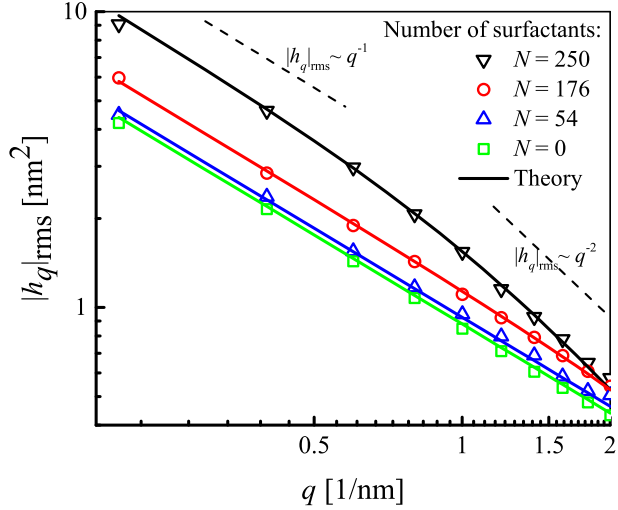}
\caption{\label{fig3} Static spectrum for capillary waves laden with different numbers ($N$) of surfactants.} 
\end{figure}

\begin{figure}[h]
\includegraphics [width=0.5\linewidth]{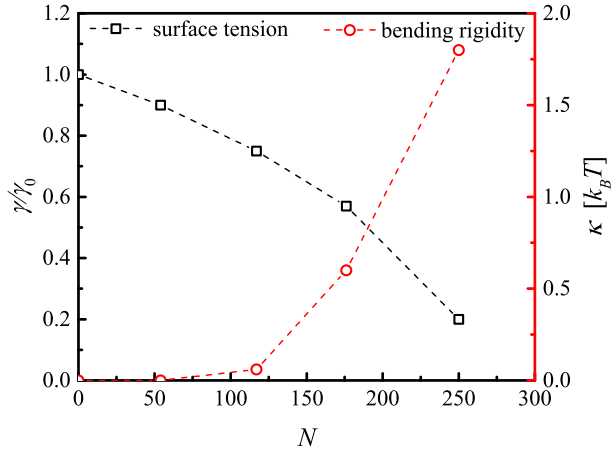}
\caption{\label{fig4} Effects of the number of surfactants on surface tension and bending rigidity. The values of surface tension are obtained by independent MD simulations while the values of bending rigidity are obtained by fitting the static spectra with predictions.}
\end{figure}

The symbols in Fig.~\ref{fig3} represent the static spectra $|h_q|_{rms}$ of surface waves obtained from simulations, which are the root mean square (rms) of surface modes $h_q(q,t)$ (averaged over 20000 times).
One can see from MD results that the wave amplitude is enhanced by increasing the number of surfactants on the liquid surface, since surfactants can decrease surface tension. The values of surface tension for the film surface contaminated with different numbers (concentrations) of surfactants can be calculated from independent MD simulations through the mechanic route\,\cite{kirkwood1949statistical} and are given in Fig.~\ref{fig4} (see the black squares). The linear relationship between surface tension and surfactant concentrations is found only be valid for low surfactant concentrations while surface tension decreases more rapidly for high surfactant concentrations.

In Fig.~\ref{fig3}, for a clean surface ($N=0$, green squares) and a surface with a low concentration of surfactants (e.g., $N=54$, blue triangles), their spectra show the conventional scaling $|h_q|_{rms}\sim q^{-1}$ for all wavenumbers, which is the scaling when only considering surface tension on the surface. Using the independently measured values of surface tension as shown in Fig.~\ref{fig4}, the MD spectra of $N=0$ and $N=54$ agree well with the theory, i.e., Eq.~(4) without the elastic term. However, when a surface is deposited with a high concentration of surfactants (e.g., $N=250$, blue triangles), the scaling changes from $-1$ to $-2$ for waves with large wavenumbers. The new scaling can be justified by the elasticity of the surfactant monolayer formed on the liquid surface, which gives an energy contribution $\sim \kappa q^4|h_q|^2$ as described by Eq.~(4). By fitting the MD results with Eq.~(4), the value of bending rigidity is obtained and its relation with surfactant concentrations is given in Fig.~\ref{fig4}, which shows that the bending rigidity increases with the number of surfactants. We note that the measurements of TCW spectra at large wavenumbers from MD simulations can depend on how to define the free surface\,\citep{delgado2008hydrodynamics}, which makes it more difficult to measure the precise value of bending rigidity if it is weak. 

The competition between elasticity and surface tension defines a length scale ${\lambda_\kappa}=2\pi\sqrt{\kappa/\gamma}$ and elasticity is dominant at lengths smaller than $\lambda_\kappa$. This can be seen from the case of $N=250$ in Fig.~\ref{fig3} where the scaling $-2$ is dominant at large wavenumbers. Using the prediction for the strength of bending rigidity $\kappa\approx k_BT$ based on the thermodynamic arguments\,\cite{helfrich1978steric}, one finds $\lambda_\kappa=2\pi\sqrt{k_BT/\gamma}$, which is on the order of thermal length and indicates that elasticity plays more important roles at the nanoscale.
 
\begin{figure}[h]
\includegraphics [width=\linewidth]{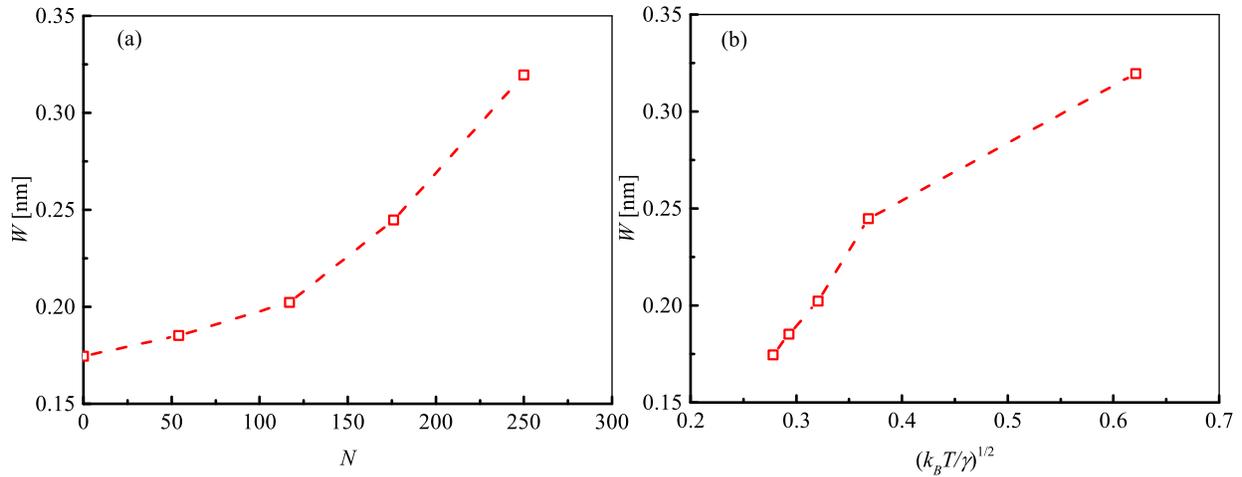}
\caption{\label{fig5} The relation of surface roughness with (a) number of surfactants and (b) thermal length.}
\end{figure}
The rms roughness of a liquid surface is calculated as (in discrete forms):
\begin{equation}
W=\sqrt{\left\langle\frac{1}{{{N}_{b}}}\sum\limits_{i=1}^{i={{N}_{b}}}{{{\left( {{h}_{i}}-\bar{h} \right)}^{2}}}\right\rangle}=\sqrt{\frac{1}{L_{x}^{2}}\sum\limits_{i=1}^{i={{N}_{b}}}{{{\left| {{h}_{q}} \right|}_{rms}^{2}}}}, 
\end{equation}
where $N_b$ is the number of bins used to extract the surface profile from MD simulations. Note that this sum of modes can be usually integrated by taking $q=2\pi/L_x$ to approach infinitely small values\,\citep{zhang2020thermal,laradji2000elastic}. However, here our film simulated is short making the integration meaningless.

 The obtained $W$ for different numbers of surfactants is shown in Fig.~\ref{fig5}(a) and it is found surface roughness is enhanced by adding surfactants. In Fig.\,\ref{fig5}(b), for $N<176$, the surface roughness is linearly proportional to thermal length $\sqrt{k_BT/\gamma}$ as predicted\,\citep{zhang2020thermal}. However, due to the strong elasticity of the liquid surface for $N>176$, this linear relationship breaks down and elasticity has the potential to reduce surface roughness, in competition with the decreased surface tension. 
\subsection{Surface viscosity of surfactant-laden liquid surface}
\begin{figure}[h]
\includegraphics [width=\linewidth]{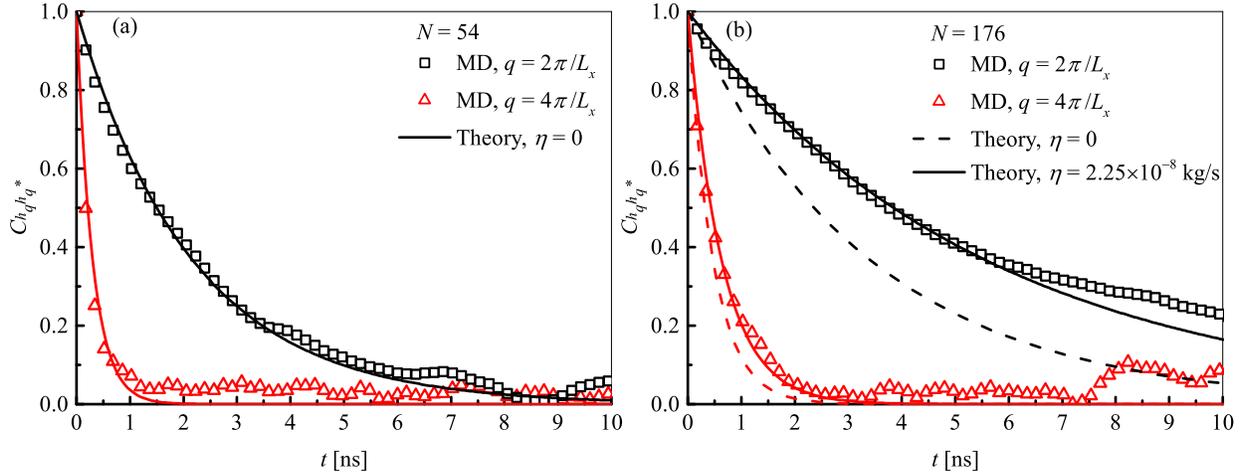}
\caption{\label{fig6} Decay of temporal correlations of surface modes at two different wavenumbers ($q=2\pi/L_x$ and $q=4\pi/L_x$) for case (a) $N=54$ and (b) $N=176$. }
\end{figure}
Fig.~\ref{fig6} shows the temporal correlations $C_{h_qh_q^*}$ of surface modes at two different wavenumbers ($q=2\pi/L_x$ and $q=4\pi/L_x$) for the case $N=54$ (see Fig.~\ref{fig6}(a)) and the case $N=176$ (see Fig.~\ref{fig6}(b)). As expected, the temporal correlations decay to zero with time and temporal correlations of waves with a larger wavenumber decay much faster\,\citep{zhang2021relaxation}.

For the case $N=54$, its MD results (represented by symbols) can be predicted nicely using Eq.~(5) with $\gamma=0.9\gamma_0$ and $\kappa=0$ independently obtained above (see Fig.~\ref{fig4}), which suggests vanishing effects of surface viscosity. However, when the number of surfactants is increased to the case $N=176$, the relaxation of correlations is significantly slowed evidenced by comparing black squares in Fig.~\ref{fig6}(b) to the black squares in Fig.~\ref{fig6}(a). Such a slow relaxation can not be explained only considering the decrease of surface tension, since the predictions (see the dash lines in Fig.\,\ref{fig6}(b)) using $\gamma=0.57\gamma_0$ and $\kappa=0.4k_BT$ (obtained above) still underestimate the MD results a lot.

\begin{figure}[h]
\includegraphics [width=\linewidth]{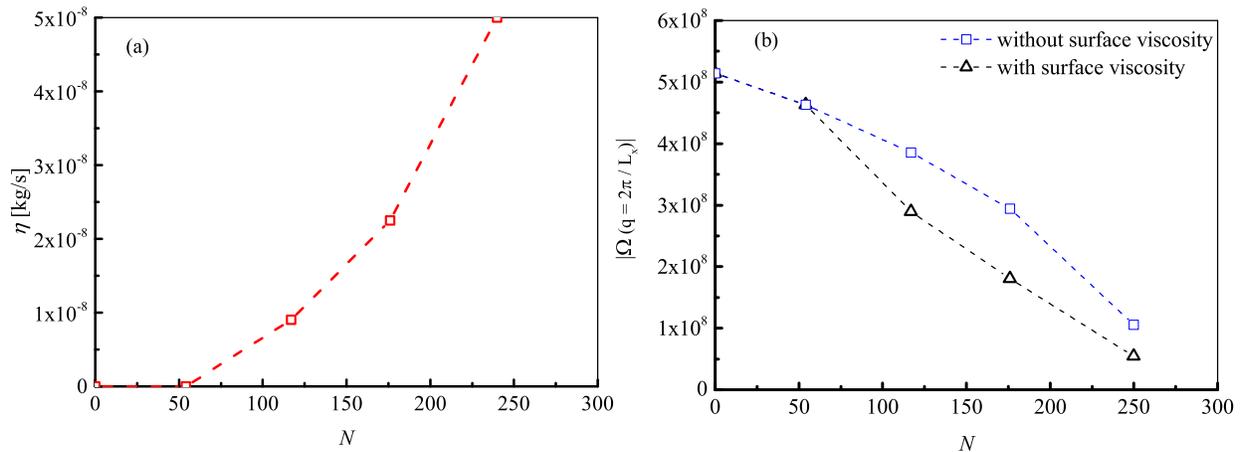}
\caption{\label{fig7}(a) The effects of the varying number of surfactants on the values of surface viscosity measured in MD. (b) The effects of the varying number of surfactants on the dispersion relation of the wave with $q=2\pi/L_x$.  }
\end{figure}

We attribute this inconsistency to surface viscosity and the MD results can be fitted by Eq.~(5) and the newly derived dispersion relation Eq.~(21) with a value of surface viscosity $\eta=2.25\times 10^{-8}$ kg/s. The inferred surface viscosity is a reasonable value compared to experimentally measured values of sodium dodecyl sulphate (SDS) solution\,\cite{stevenson2005remarks}, which ranges from $10^{-8}$ to $10^{-6}$ kg/s in different measurement techniques. 
 By fitting temporal correlations with Eq.~(5) and Eq.~(21) for other surfactant concentrations, we obtain the relation between surface viscosity and the number of surfactants shown in Fig.~\ref{fig7}(a), which shows the same trend as experimental results\,\citep{choudhury2020enhanced,tambe1994effect}. Fig.~\ref{fig7}(b) shows the reduction of the dispersion relation (for $q=2\pi/L_x$) by increasing the number of surfactants at the free surface. The reduction is due to both the decreased surface tension (blue line) and the increased viscosity(black line).

\begin{figure}
\includegraphics [width=\linewidth]{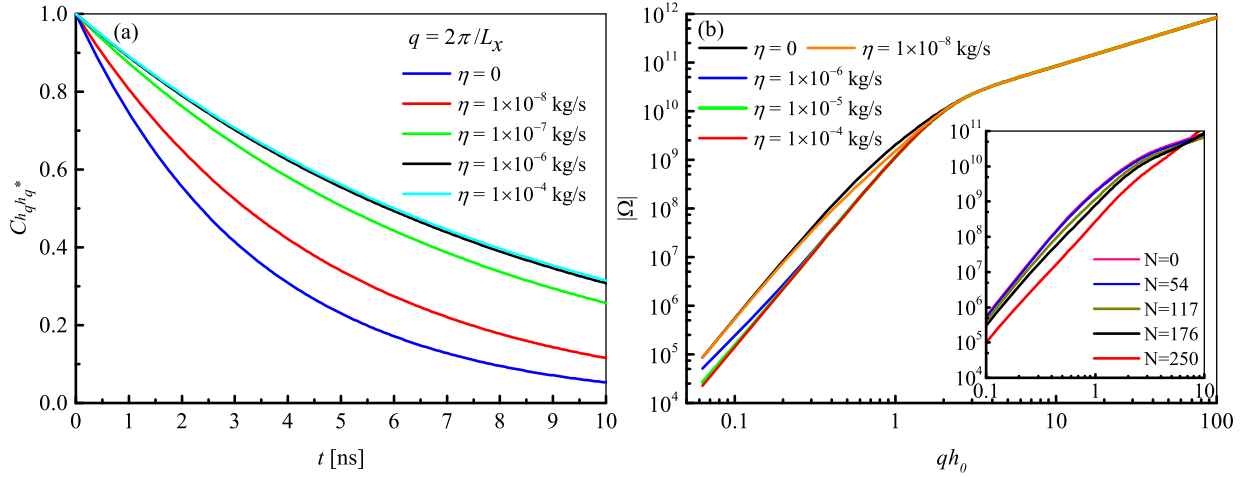}
\caption{\label{fig8} (a) Damping effects of surface viscosity on the relaxation of correlations. The other parameters used are $\gamma=0.57\gamma_0$ and $\kappa=0.4k_BT$. (b) Effects of surface viscosity on the dispersion relation with $\gamma=\gamma_0$ and $\kappa=0$. The inset shows the dispersion relation with a different number of surfactants.}
\end{figure}

The ability of surface viscosity to slow relaxations of temporal correlations has its maximum, as plotted in Fig.~\ref{fig8}(a), where a further increase of surface viscosity beyond $\eta=1\times 10^{-6}$ kg/s leads to very small changes of the correlations.   
Simplifications of the newly derived dispersion relation can be done for two asymptotic limits $qh_0\ll 1$ and $qh_0\gg 1$ (Notably $qh_0 =0.63$ in our work for $q=2\pi/L_x$) as shown in Fig.~\ref{fig8}(b). In the limits of $qh_0\ll 1$ and no slip, the dispersion relation is simplified to 
\begin{equation}
\Omega =-\frac{\left( \gamma {{q}^{2}}+\kappa {{q}^{4}} \right){{q}^{2}}h_{0}^{3}}{12\mu }\frac{4\mu+{{h}_{0}}\eta {{q}^{2}}}{\mu+{{h}_{0}}\eta {{q}^{2}}},
\end{equation}
which is the dispersion relation for thin liquid films obtained earlier by Edwards and Oron\,\cite{edwards1995instability} though they did not have the elastic term in Eq.~(25). In the limits of zero surface viscosity, the constant factor in Eq.~(25) is $1/3$ (suggesting a no-shear boundary condition at the free surface) while it is $1/12$ (suggesting a tangentially immobile condition at the free surface) in the limits of infinite surface viscosity. Thus, surface viscosity has at most a fourfold ability to decrease the relaxation of waves with $qh_0\ll 1$, as demonstrated by the dispersion relation in Fig.~\ref{fig8}(b). In our simulations, due to the reduced surface tension, the decrease of dispersion relation by increasing the number of surfactants is larger than four times  (see Fig.~\ref{fig7}(b) and the inset of Fig.~\ref{fig8}(b) for the case $N=250$).

However, in the opposite limits $qh_0\gg 1$, Eq.~\eqref{eq:disp_rela} is simplified to
\begin{equation}
\Omega =-\frac{ \gamma +\kappa q^2 }{2\mu }q,
\end{equation}
which is independent of surface viscosity. This is can be seen from Fig.~\ref{fig8}(b) where all solid lines with different surface viscosity overlap each other for wavenumbers $qh_0\gg 1$. Notably, this counter-intuitive result is due to the assumption of negligible inertia terms in momentum equations, which is usually valid at the nanoscale. For capillary waves on macroscopic films with infinite depth ($qh_0\gg 1$), Shen \textit{et al.}\,\cite{shen2018capillary} considered both the effects of surface viscosity and inertia and derived a dispersion relation depending on surface viscosity. Equation (26) in our work can be recovered by neglecting the inertia terms of the dispersion relation in Ref.\,\cite{shen2018capillary}.

\begin{figure}[h]
\includegraphics [width=0.5\linewidth]{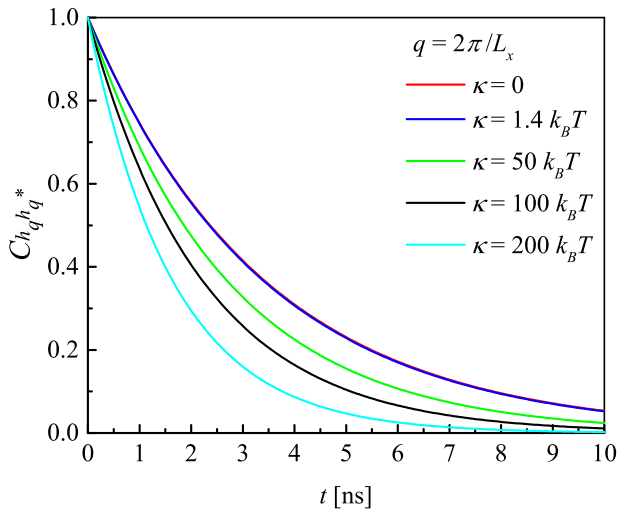}
\caption{\label{fig9} The effects of elasticity on the relaxation of temporal correlations. The parameters used are $\gamma=0.57\gamma_0$ and $\eta=0$.}
\end{figure}

Finally, we remark that the bending rigidity has the potential to speed up the relaxation of correlations but in our simulations its values are too weak to have noticeable effects on the wavenumber $q=2\pi/L_x$ (see the red line with $\kappa=0$ and the blue line $\kappa=1.4 k_B T$ in Fig.~\ref{fig9}). However much larger bending rigidity can be obtained in experiments\,\cite{dressaire2008interfacial} so that the effects of elasticity can be significant (e.g. the black line with $100k_BT$ in Fig.~\ref{fig9} ).  

\section{Conclusions}\label{sec5}
In this work, the effects of surfactants on the statics and dynamics of thermal capillary waves are studied both theoretically and numerically. Molecular simulations are used to simulate surfactant-laden free surface flows. The obtained static spectra show surfactants can enhance the elasticity of the liquid surface apart from reducing surface tension. The slower decay of temporal correlations of surface modes is observed in simulations by adding surfactants and can be predicted by a newly derived dispersion relation considering surface viscosity. We believe the new theories developed here and molecular simulations performed can motivate new experiments using thermal capillary waves to measure surface viscosity. 

It is worth mentioning that our current molecular system is ideal, in the future, it is interesting to use more realistic molecular models or map our LJ system to water or other simple globular solvents in colloidal science. We also remark that the choice of the model for surfactants or parameters in the FENE model adopted here is not unique. For example, Laradji and Mouritsen\,\citep{laradji2000elastic} use a different model for the surfactant. The original FENE model\,\citep{kremer1990dynamics} use the same cutoff for all beads unlike us. In the future, a systematic test of models and their parameters is needed to see how they change the elasticity and surface viscosity.

Thermal fluctuations of fluid surface have been shown to increase the instability of nanofilms and nanojets\,\cite{zhang2019,zhao2019revisiting}. As surfactants can decrease surface tension and increase thermal roughness, the presence of surfactants may enhance the instability and rupture of nanofilms. For nanobubbles, one may intuitively think surfactants can stabilize surface nanobubbles by reducing surface tension and the pressure-driven diffusion of gas from bubbles. However, experiments have shown the opposite effect of surfactants\,\cite{lohse2015surface,zhang2012effects} and this may be due to the instability of the bubble surface caused by extremely small surface tension and large surface fluctuations. 
\section{Acknowledgments}
Zhang wishes to thank for the discussions with Detlef Lohse, Yibo Chen and Hongguang Zhang.
Ding wishes to thank the financial support from National Natural Science Foundation of China (Grant No. 12102109).
\section*{appendix: Distribution of surfactants}
\renewcommand{\theequation}{A\arabic{equation}}
\setcounter{equation}{0}
\renewcommand{\thefigure}{A\arabic{figure}}
\setcounter{figure}{0}
\begin{figure}[h]
\includegraphics[width=\linewidth]{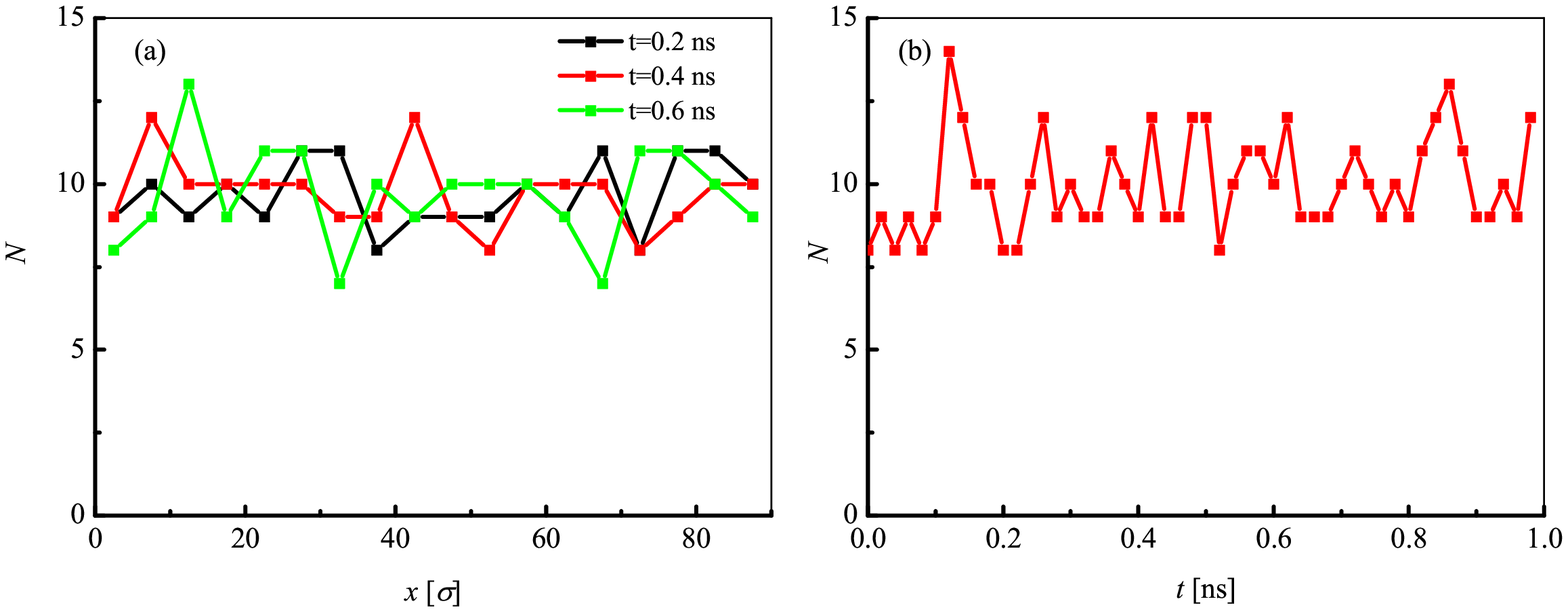}
\caption{(a) Distribution of the number of surfactants along the surface. (b) Change of the number of surfactants with time in the middle region of the surface.}  \label{figA1}
\end{figure}

Fig.~\ref{figA1}(a) shows the distribution of surfactants along the surface at three consecutive time while Fig.~\ref{figA1}(b) shows the change of the number of surfactants in the middle region of the surface. Though the number of surfactants fluctuates with time due to the nature of molecules, it can be seen that the distribution of surfactants is uniform along the surface. 
%
\end{document}